\documentclass[journal=jpclcd,manuscript=letter]{achemso}

\usepackage{graphicx}
\usepackage{dcolumn}
\usepackage{bm}
\usepackage{hyperref}
\usepackage{array} 
\usepackage[export]{adjustbox}
\usepackage[nomessages]{fp}
\usepackage{xcolor}
\usepackage{placeins}
\usepackage[version=3]{mhchem} 
\usepackage[T1]{fontenc}       

\author{Kun Yao}
\author{John E. Herr}
\author{Seth N. Brown}
\author{John Parkhill}%
\email{jparkhil@nd.edu}
\affiliation{%
 Dept. of Chemistry and Biochemistry, The University of Notre Dame du Lac 
}%
\title[An \textsf{achemso} demo]
  {Intrinsic Bond Energies from a Diatomics-in-Molecules Neural Network}

\begin{document}

\begin{abstract}
Neural networks are being used to make new types of empirical chemical models as inexpensive as force fields, but with accuracy similar to the \emph{ab-initio} methods used to build them. Besides modeling potential energy surfaces, neural networks can provide qualitative insights and make qualitative chemical trends quantitatively predictable. In this work we present a neural network that predicts the energies of molecules as a sum of intrinsic bond energies. The network learns the total energies of the popular GDB9 dataset to a competitive MAE of 0.94 kcal/mol on molecules outside of its training set, is naturally linearly scaling, and applicable to molecules of consisting of thousands of bonds. More importantly it gives chemical insight into the relative strengths of bonds as a function of their molecular environment, despite only being trained on total energy information. We show that the network makes predictions of relative bond strengths in good agreement with measured trends and human predictions. A Diatomics-in-Molecules Neural Network (DIM-NN) learns heuristic relative bond strengths like expert synthetic chemists, and compares well with ab-initio bond order measures such as NBO analysis. 
\begin{tocentry}
\includegraphics[width=1.0\textwidth]{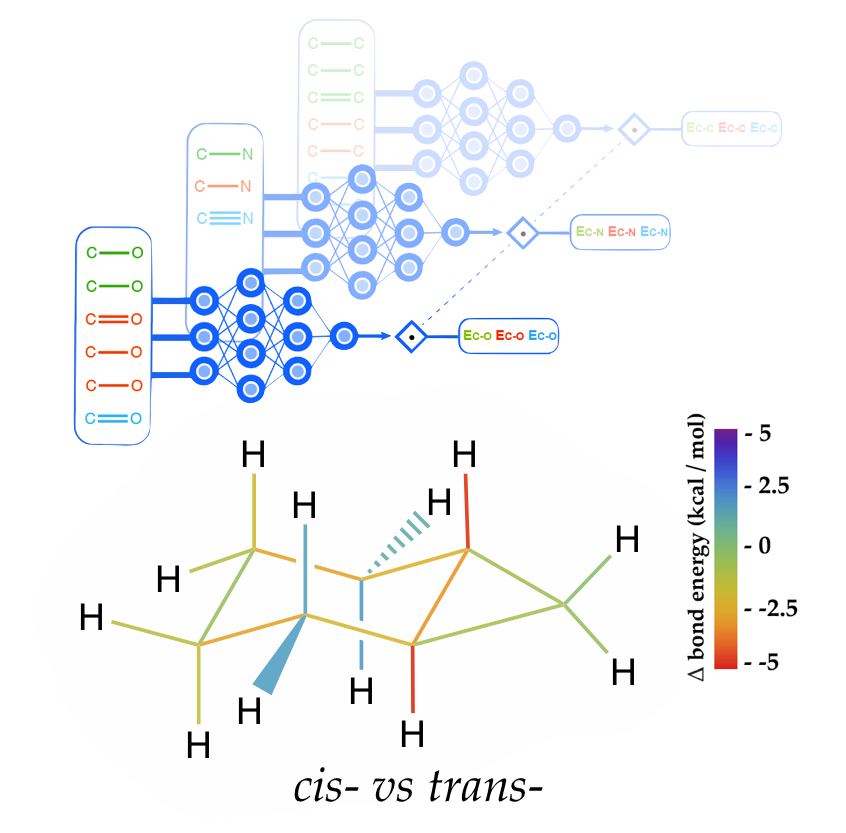}
\end{tocentry}

\end{abstract}


\indent Neural networks (NN) make accurate models of high dimensional functions, with chemical applications such as density functionals\cite{Snyder:2013aa, snyder2012finding, yao2016kinetic}, potential energy surfaces\cite{behler2011neural, handley2010potential, schutt2016naturecm,chmiela2016machine,tian2016hierarchical, behler2007generalized,behler2011atomcentered, gastegger2016comparing, morawietz2016van, bartok2010gaussian, mones2016exploration, khaliullin2011nucleation, smith2017ani, jiang2013permutation,zhang2014effects, shao2016communication, li2015permutationally, li2015permutationally2, kolb2017high,deringer2016machine, medders2015representation, medders2013critical, conte2015permutationally, manzhos2006random, manzhos2014neural, manzhos2009fitting, malshe2010input} and molecular properties\cite{CM,BoBs,Hansen:2015aa,montavon2013machine, pilania2013accelerating, ghasemi2015interatomic, schutt2014represent, olivares2011accelerated, ma2015machine, ediz2009using, lopez2014modeling, cuny2016ab, hachmann2011harvard, hachmann2014lead, simon1993combined, hautier2010finding, ediz2008molecular, barker2016localized,chen1997knowledge,  collins2017constant, wu2017moleculenet, isayev2015materials, rupakheti2015strategy, bodor1991neural, virshup2013stochastic,qu2013big, nelson2012nonadiabatic,kvasnicka1993neural}. With modern general-purpose graphical processing unit (GP-GPU) computing, NN's are inexpensive to train and evaluate, with a cost that lies much closer to a force-field than ab-initio theory. To model energy extensively, most black-box NN schemes partition the energy of a molecule into fragments\cite{behler2007generalized, bartok2010gaussian, medders2013critical, yao2017many}. Separate networks are often used for qualitatively different energy contributions to maximize accuracy and efficiency. Besides balancing accuracy with complexity, a decomposition of the energy can yield chemical insights, and give the heuristic principles of chemistry reproducible, quantitative rigor. In this paper we use a NN to express the cohesive energy of a molecule as a sum of bond energies. Besides offering a very accurate and inexpensive decomposition of the total energy, this method, Diatomics-in-Molecules Neural Network (DIM-NN) produces an instant estimate of the embedded unrelaxed bond strengths in a molecule. DIM-NN responds quantitatively to molecular geometry in a way that textbook tabulated values cannot, and updates a textbook concept used by all chemists making it a quantative tool.\\
\indent     Atoms have been a popular choice of NN energy decomposition since the pioneering contributions of Behler, Khaliullin and Parinello\cite{behler2007generalized,behler2011atomcentered, behler2011neural, khaliullin2011nucleation}. Within the atom scheme separate networks are trained for each element. Another reasonable choice for non-covalent aggregates explored in our own work is a many-body expansion\cite{yao2017many}. These different fragmentations of the energy navigate a trade-off between breadth and accuracy. One atom's contribution to the total energy is difficult to learn because it varies significantly when engaged in different bonds. One sort of three-body interaction \cite{medders2013critical, yao2017many, manzhos2014neural}is comparatively easy to learn, but there are too many three-body combinations in chemistry to learn them all. The purpose of this paper is to explore the advantages of bonds as a decomposition unit. \\
\begin{figure*}
\includegraphics[width=1.0\textwidth]{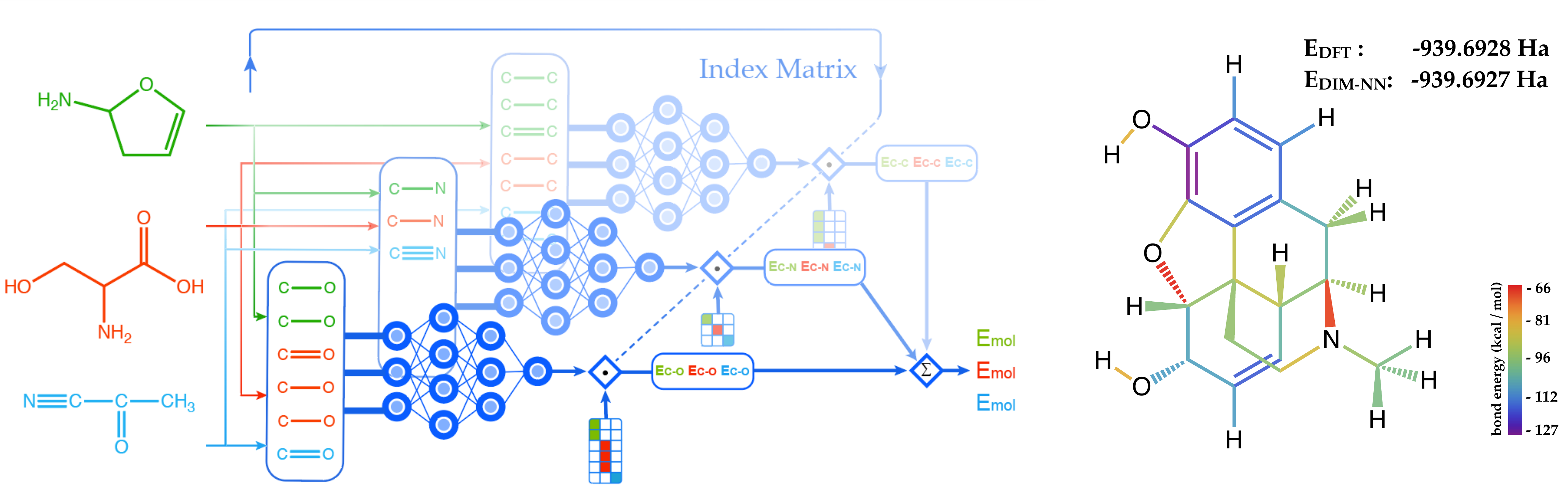}
\caption{Left panel: A scheme of how DIM-NN computational graph is trained. A batch of molecules is shattered into diatomics. Bond types within the batch are fed into a type-specific sub-network and a linear transformation reassembles molecular energies. Right panel:Morphine, a molecule not included in the training set, with its bonds color-coded by their bond energies as calculated using DIM-NN. The total energy as a sum of bonds is accurate within 1mE$_h$.}
\label{Fig1}
\end{figure*}
\indent     Bonds have served as the central abstraction throughout the history of chemistry\cite{tully1973diatomics1, tully1973diatomics2}. They are a popular building block for Neural Network inputs\cite{BoBs}, but to our knowledge no NN has been used to predict total energies as a sum of bond energies as we are about to describe. One reason for this might be the large, but manageable, number of bond networks required to make a general model chemistry. A complex neural network is required for DIM-NN with many bond-branches that must be dynamically learned and evaluated. We developed a general open-source software framework for producing NN models\cite{tensorflow2015-whitepaper} of molecules, TensorMol in which we have implemented DIM-NN, that simplifies the process of creating the bond-centered network. The complete source allowing readers to reproduce and extend this work is publicly available in the TensorMol repository\cite{TENSORMOL}.\\
\indent     The left panel of Fig. \ref{Fig1} schematically describes how DIM-NN is trained and evaluated. A molecule is broken down into overlapping diatomic fragments, such as C-H, C-C, C-O, etc. An optimal choice of descriptor describing the chemical environment of the bond is crucial for neural networks to reach their best performance\cite{behler2011atomcentered, bartok2013representing, jiang2013permutation, CM, faber2017fast, collins2017constant, faber2017fast}. This work uses our own version of the Bag of Bonds (BoBs)\cite{BoBs, CM} descriptor, which is similar to BALM \cite{BAML} for this purpose. The descriptor contains the bond length of the target bond, and lengths and angles of bonds attached to it in order of connectivity. Each type of bond has its own branch consisting of three fully-connected hidden layers with 1000 neurons in each layer summed to the bond energy. The energies of all bonds are summed up at the last layer to give a predicted molecular total energy, which is the training target. Backpropagation of errors\cite{Rumelhart:1986fk} to the previous layers allows the bond branches to learn the strengths of their specific bond types. These errors propagate back through a linear transformation matrix that maps the many bond energies coming from different molecules produced by a type branch back onto the molecules.  We use the popular GDB9 database\cite{gdb9, virshup2013stochastic} including H, C, N and O for our training data, which is more than 130,000 molecules in total. Before training, 20\%\ percent of the database was chosen randomly as a test set and kept independent. Our results also examine larger molecules, which are not a part of GDB9.  Further methodological details are presented in the supplement.\\
\begin{figure}[t]
\includegraphics[width=0.7\textwidth]{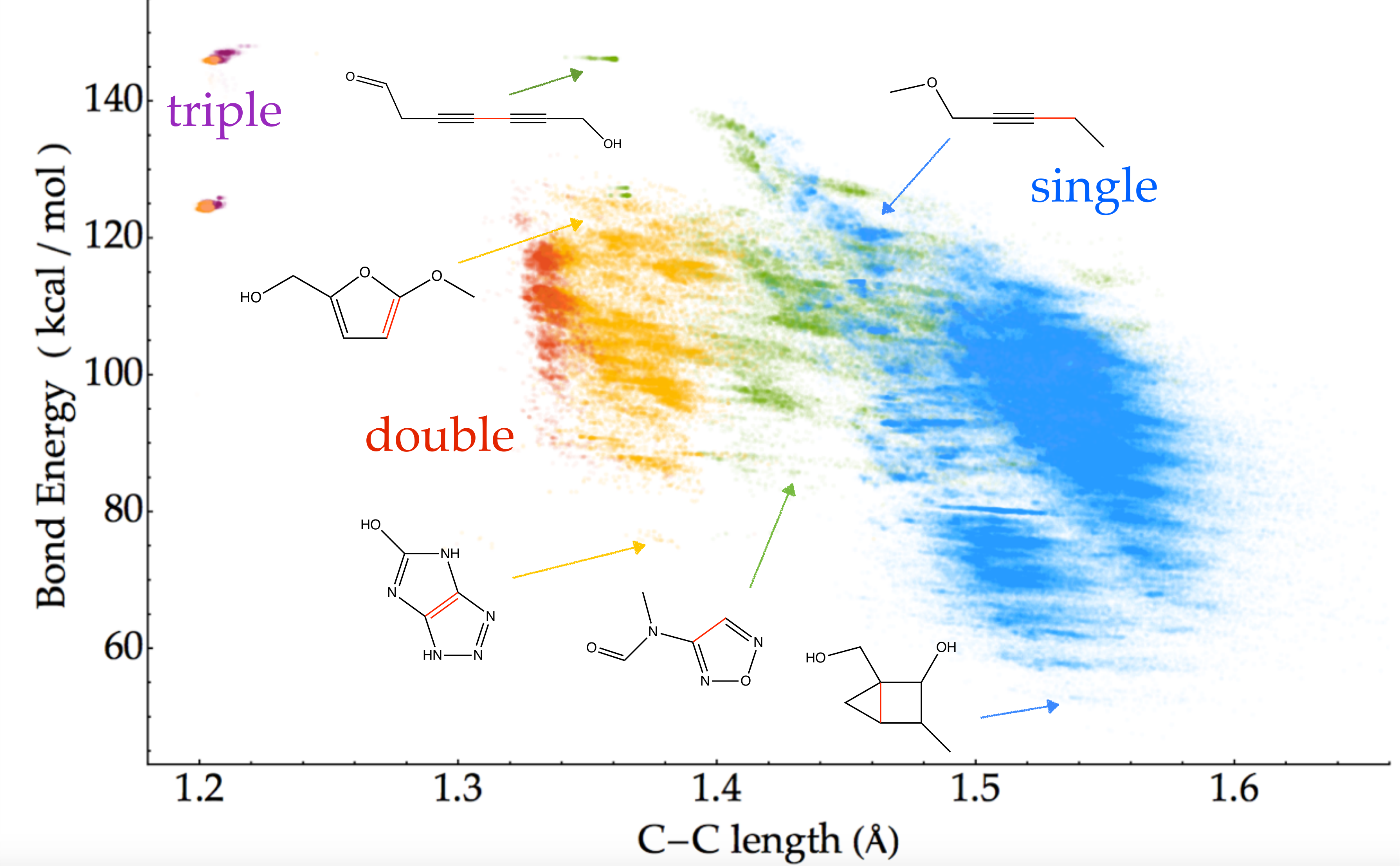}
\caption{ DIM-NN predicted bond energies of the 720,000 C-C bonds in GDB9 plotted as a function of bond length. Single, conjugated single, double, conjugated double, triple and conjugated triple bonds are shown in blue, green, red, yellow, orange and purple, respectively. Exemplary strong and weak bonds are annotated. The roundness of each region shows that bond length is roughly as significant as all the other factors which modulate the strength of bonds, combined.}
\label{Fig2}
\end{figure}

\indent Our results try to answer three questions we had about DIM-NN: how well does DIM-NN predict total energies, how well does DIM-NN predict bond strength trends, and what is the relationship between DIM-NN bond energies and observed chemical reactivity. We were especially excited about DIM-NN as a method to systematically and quantitatively reproduce chemical heuristics, and appealed to expert synthetic chemists to make predictions of relative bond strengths to test DIM-NN. Because DIM-NN predicts the energy of a molecule relative to free neutral atoms, we expect bond strengths in DIM-NN to correlate best with unrelaxed, heterolytic dissociation energies and some evidence to this effect is presented later on. However, no specific products are implied by DIM-NN bond energies. They depend only on the equilibrium geometry. Some sorts of bonds are strong near equilibrium but can kinetically access their products which are relatively more stable, or electronically relax, and in these cases a chemist's intuition may be at variance with DIM-NN.\\
\indent     The training mean absolute error (vs. B3LYP/6-31G(2df,p) ) of the molecular energies of our DIM-NN reaches 0.86 kcal/mol and the independent test mean absolute error is 0.94 kcal/mol. The test error is close to the training error, suggesting that the our model is neither over-fitting nor under-fitting.  The accuracy of our model is competitive with the state-of-art sub-1 kcal/mol accuracy on this dataset.\cite{BoBs, BAML, collins2017constant, barker2016localized, hansen2013assessment,schutt2016naturecm, faber2017fast,kearnes2016molecular}. The bond-wise nature of our model makes it transferable to large molecules. The right panel of Fig. \ref{Fig1} shows DIM-NN predicted total energy and the calculated DFT total energy of morphine molecule, which is not in our training set and contains 21 heavy atoms. We also tested DIM-NN on vitamins D2 and B5. The difference between DIM-NN and DFT energies of these two molecules are 0.6 kcal/mol and 1.2 kcal/mol, respectively. All these errors are small relative to the inherent errors of the B3LYP model chemistry used to produce DIM-NN. Our model can be trivially trained on higher quality chemical data as it becomes available. Correct reproduction of the bond trends described later on depends sensitively on the accuracies of the total energies. Before completion of the training process when total energies are in error by roughly 5 kcal/mol, more than 40\% of the examples in Table 1 and Table S1 are answered incorrectly by DIM-NN. This is due to the fact that errors can accumulate in bonds and cancel in the molecular energy unless the NN is tightly trained on a broad sample of chemical space.\\

\begin{figure}[t]
\includegraphics[width=0.8\textwidth]{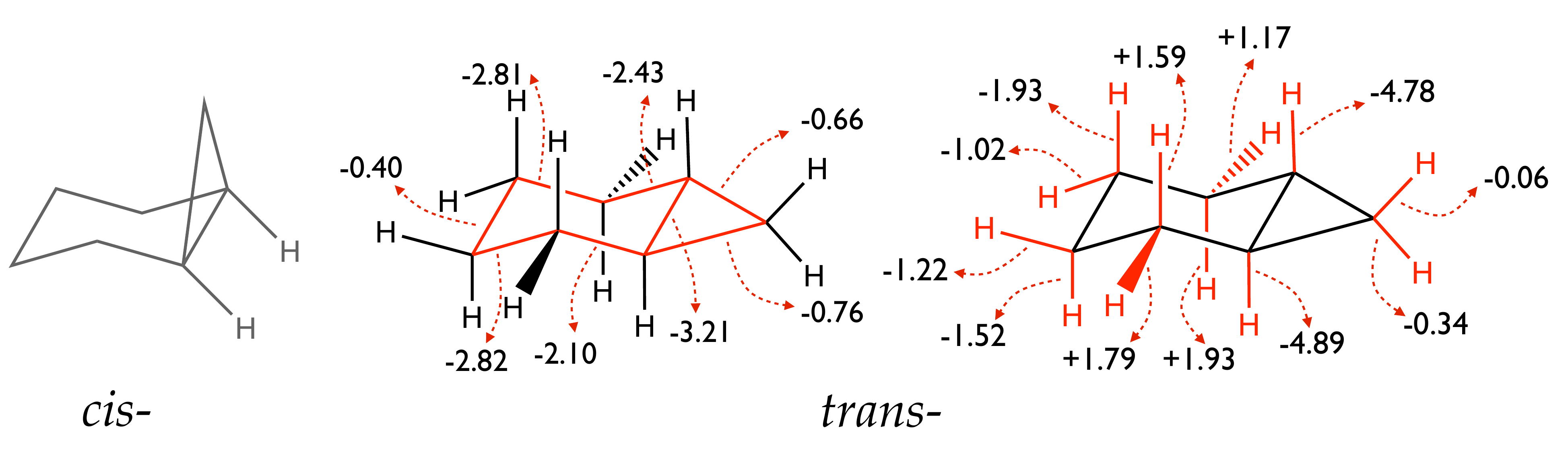}
\caption{Bond energies (kcal/mol) of \textit{trans-}bicyclo[4.1.0]heptane, given as changes relative to the \textit{cis-}isomer show how strain energy is distributed mostly over the cyclohexane ring.}
\label{Fig3}
\end{figure}

\indent  Our scheme not only calculates the total energy of a molecule, but also predicts the strengths of bonds individually. Fig. \ref{Fig1} shows the bonds in the morphine molecule drawn with colors keyed to the bond strengths predicted by DIM-NN. The slight energy differences predicted by DIM-NN between bonds with identical bonding but different environment are perceptible. DIM-NN can suggest how the stress is distributed in an unstable isomer compared with the stable one. Fig. \ref{Fig3} shows the geometry of \textit{cis-} and \textit{trans-}bicyclo[4.1.0]heptane. The DFT calculation shows that \textit{cis-} structure is 26.3 kcal/mol more stable than the \textit{trans-}  structure. DIM-NN predicts the energy difference is 24.6 kcal/mol, in good agreement for this molecule outside GDB9. Fig. \ref{Fig3} also shows how the stress is distributed in the \textit{trans-} structure. Within the carbon framework, the strain is distributed rather equally among the bonds in the cyclohexane ring. Some C-H bonds become weaker, especially those attached to at the ring junction carbons, and some become stronger. \\
\indent Fig. \ref{Fig2} shows the bond energies of all the carbon-carbon bonds in the GDB9 database with respect to their bond lengths. Different bond types (single, double, triple and conjugated versions of those) are indicated by color. The spread of each clouds is caused by the different chemical environments of the bonds. Bond strength correlates predictably with bond order, but general chemical trends which are less obvious also revealed. For example energies of single bonds are more sensitive to the environment than those of double or triple bonds. DIM-NN also predicts that bond lengths are roughly as relevant to their energies as all other factors combined.\\
\indent     DIM-NN predicts that a bi-cyclic C-C single bond shared by a three-membered carbon ring and a four-membered carbon ring is extremely weak, consistent with chemical intuition. The strongest C-C single bond predicted by DIM-NN is the bond that is connected with C-C triple bonds, in agreement with textbook bond dissociation energies\cite{mcmillen1982hydrocarbon, speight2005lange}. DIM-NN bond energies reproduce several other established chemical rules of thumb, for example that the strength of a CH bond decreases in the series:  methyl carbon, primary carbon in ethane, secondary carbon in propane, tertiary carbon in isobutane. The bond energies of these four types of carbon-hydrogen bonds predicted by DIM-NN are 105.2 kcal/mol, 105.0 kcal/mol, 95.2 kcal/mol and 89.8 kcal/mol respectively, where the experimental value are 105.1 kcal/mol, 98.2 kcal/mol, 95.1 kcal/mol and 93.2 kcal/mol, respectively\cite{mcmillen1982hydrocarbon}. DIM-NN also predicts that the C-C single bond in pyrrole is 1.2 kcal/mol more stable than the C-C single bond in furan, which agrees with greater bond delocalization in pyrrole than in furan, consistent with its larger NICS aromaticity\cite{horner2013chemical,Schleyer:1996aa}. \\

\begin{table*}
  \centering
  \begin{tabular}{ c  m{6cm}  c   c   c  }
    \hline
    Case \# &  & Chemist & Neural Network & NBO \\ \hline 
    1 &\begin{minipage}{.5\textwidth}
    \adjustbox{right=4.5cm}{\includegraphics[height=1.2cm]{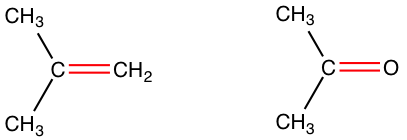}}
    \end{minipage}
    & $\ll$ &  \FPeval{\result}{round((0.161473-0.226009)*627.503, 1)}%
    $\result$  & \FPeval{\result}{round(3.95128-3.98066,4)}%
    $\result$ \\ \hline
    2 & \begin{minipage}{.5\textwidth}
     \adjustbox{right=4.6cm}{\includegraphics[height=1.2cm]{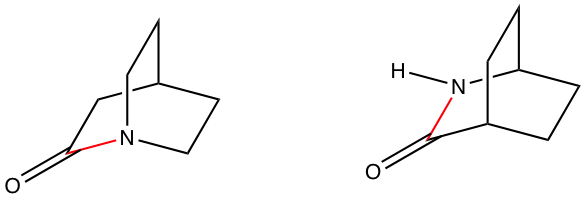}}
    \end{minipage}
    & $\ll$ & \FPeval{\result}{round((0.123871-0.15127)*627.503, 1)}%
    $\result$ & \FPeval{\result}{round(1.98041 - 1.99005,4)}%
    $\result$ \\ \hline
    3 & \begin{minipage}{.5\textwidth}
    \adjustbox{right=5.5cm}{\includegraphics[height=1.2cm]{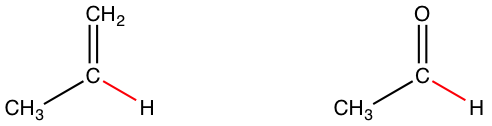}}
    \end{minipage}
    & $\gg$ & \FPeval{\result}{round((0.16994-0.196219)*627.503, 1)}%
    $\color{red} \result$ & \FPeval{\result}{round(1.97834- 1.98866,4)}%
    $\color{red} \result$ \\ \hline
    4 & \begin{minipage}{.5\textwidth}
     \adjustbox{right=4.8cm}{\includegraphics[height=1.2cm]{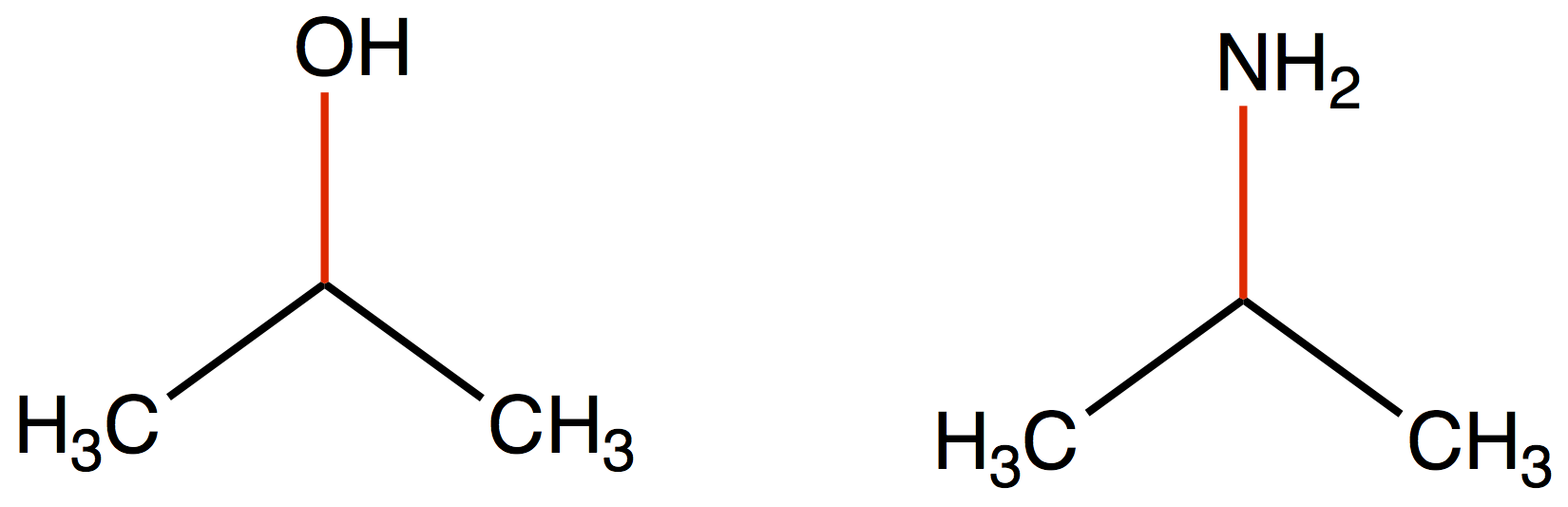}}
    \end{minipage}
    & $>$ & \FPeval{\result}{round((0.167324-0.144648)*627.503, 1)}%
    $\result$ & \FPeval{\result}{round(1.99311-1.99116,4)}%
    $ \result$ \\ \hline
    5 & \begin{minipage}{.5\textwidth}
    \adjustbox{right=6.0cm}{\includegraphics[height=1.2cm]{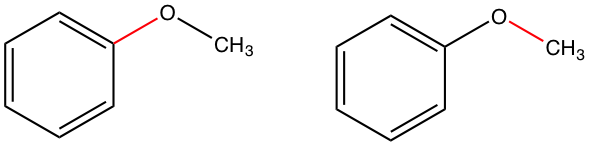}}
    \end{minipage}
    & $>$ & \FPeval{\result}{round((0.139287-0.122267)*627.503, 1)}%
    $\result$ & \FPeval{\result}{round(1.99121-1.99278,4)}%
    $\color{red} \result$ \\ \hline
    6 & \begin{minipage}{.5\textwidth}
    \adjustbox{right=5cm}{\includegraphics[height=1.2cm]{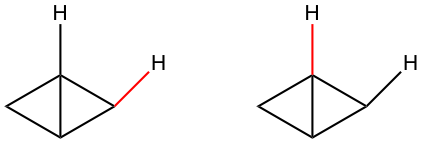}}
    \end{minipage}
    & $<$ & \FPeval{\result}{round((0.172799-0.163148)*627.503, 1)}%
    $\result$ & \FPeval{\result}{round((1.98846+1.99331)/2.0-1.99392,4)}%
    $\result$ \\ \hline
    7 & \begin{minipage}{.5\textwidth}
    \adjustbox{right=5.4cm}{\includegraphics[height=1.2cm]{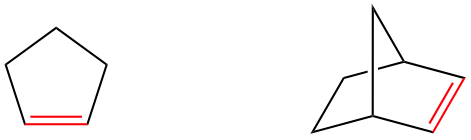}}
    \end{minipage}
    & $\gg$ & \FPeval{\result}{round((0.178594-0.172557)*627.503, 1)}%
    $\result$ & \FPeval{\result}{round(3.94474-3.94233,4)}%
    $\result$ \\ \hline
    8 & \begin{minipage}{.5\textwidth}
    \adjustbox{right=4.5cm}{\includegraphics[height=1.2cm]{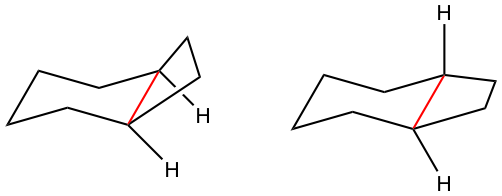}}
    \end{minipage}
    & $>$ & \FPeval{\result}{round((0.128708-0.123976)*627.503, 1)}%
    $\result$ & \FPeval{\result}{round(1.97334-1.96634,4)}%
    $\result$ \\ \hline
    9 & \begin{minipage}{.5\textwidth}
    \adjustbox{right=6.2cm}{\includegraphics[height=1.2cm]{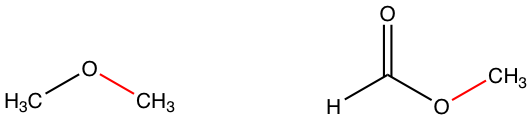}}
    \end{minipage}
    & $>$ & \FPeval{\result}{round((0.123928-0.121061 )*627.503, 1)}%
    $\result$ & \FPeval{\result}{round(1.993764-1.99221,4)}%
    $\result$ \\ \hline
    10 & \begin{minipage}{.5\textwidth}
    \adjustbox{right=4.8cm}{\includegraphics[height=1.2cm]{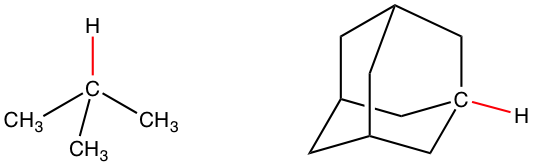}}
    \end{minipage}
    & $<$ & \FPeval{\result}{round((0.143126-0.145769)*627.503, 1)}%
    $\result$  & \FPeval{\result}{round(1.97038-1.97606,4)}%
    $\result$ \\ \hline
  \end{tabular}
  \caption{Relative strengths of the bond highlighted in red, as predicted by a chemist's intuition, DIM-NN and NBO analysis. $\gg$ and $\ll$ means the chemist expected the difference to be larger than 10 kcal/mol, $>$ and $<$ means the difference to be between 1 $\sim$ 10 kcal/mol. The DIM-NN column shows the predicted relative bond strength in kcal/mol. The relative bond strength is calculated by take the bond energy of left-hand bond and subtracting the bond energy of the right-hand bond. The number in the NBO column is the difference of the occupation number predicted by NBO.}\label{tbl:myLboro}
\end{table*}

\indent    We asked a synthetic colleague for a quiz which consists of 19 pairs of bond strength comparisons. We compare the predictions of relative bond strength made by the chemist, natural bond orbital analysis (NBO)\cite{reed1985natural,foster1980natural, weinhold2012natural} and DIM-NN. Table 1 and Table S1 shows that the problems in this quiz range in difficulty from pairs separated by $>$10 kcal/mol to subtle differences on the order of $k_bT$ which challenge the density functional data used to produce DIM-NN. NBO makes predictions disagreeing with the chemist in five cases, while DIM-NN disagrees with chemist on two cases: case 3 and case S6. Both of these two cases are comparisons of C-H bond strength where the carbon atom is connected to an oxygen atom and NBO also dissents with the chemist in these cases. We believe this disagreement is due to the fact that both NBO and the DIM-NN scheme do not relax the electronic structure of a molecule following bond cleavage, while a chemist takes into account the stabilization of the dissociated radicals. \\
\indent To further corroborate the interpretation of DIM-NN bond energies as unrelaxed homolytic bond dissociation energies (BDEs), we have directly compared DIM-NN bond energies with experiment and $\Delta$DFT. The mean absolute difference of BDEs relative to experimental best estimates from DIM-NN, geometrically unrelaxed, and relaxed DFT are 12.8, 17.2, and 8.3 kcal/mol respectively (Table S2).  The unrelaxed DFT BDEs use the geometry of the molecule, but relax the electronic wavefunction. Please note that by design DIM-NN does not include either electronic or geometric relaxation energy. The fact that DIM-NN outperforms unrelaxed DFT may indicate a cancellation of electronic and geometric relaxation energies. \\
\indent     DIM-NN bond energies are very close to experimental BDEs for simple alkanes which do not undergo significant electronic or geometric relaxation (methyl, cyclopentyl, ethyl, cyclopropyl, t-butyl, $H-CH(CH_{3})CN$, etc.), even closer to experiment than DFT. The DIM-NN bond energy is systematically larger than the experimental BDE when the product radical is electronically stabilized (allyl, tolyl, cyclopentadienyl, 1,4 cyclohexyldienyl, and $C-H$ bonds $\alpha$ to oxygen). The difference in these cases can be interpreted as the electronic relaxation energy of the radical when the difference between relaxed and unrelaxed DFT is small.\\
\indent     To investigate the electronic relaxation effect we performed embedded DFT calculations of the relative CH BDEs in the tetrahydrofuran and actealdehyde-propene examples from Table 1, and estimated the relaxation effect by comparing the difference of the fragment reaction energies frozen at their electronic configuration in the molecule\cite{manby2012simple}. In both cases the frozen-embedded calculation predicts the same ordering as DIM-NN and NBO. DIM-NN successfully learns many important classes of chemical heuristics such as bond types (case 1, case 4), geometric stresses (case 7, case 8) and conjugation (case 2). The interpretation consistent with these results is that DIM-NN produces intrinsic bond energies, which can be used to separate the stabilization of products from the intrinsic stabilization of a bond.\\
\indent We have presented a method to cheaply and accurately sum-up a molecular total energy as a ensemble of bond-energies, DIM-NN. The method can be thought of as a quantitative version of the textbook bond energy table all chemists think about and rely upon to understand molecules. Chemists could use DIM-NN to study how small changes in geometry might affect the strength of a bond, and produce quantitative numbers to match their qualitative intuition. The method also produces accurate total energies without any sort of sophisticated non-local interaction between the bonds. This shows that bond networks could contribute to a general neural network model chemistry. This ambitious long term goal merits future work, for example extending our bond decomposition with additional non-covalent contributions that describe correct long-range forces\cite{tkatchenko2012accurate}. Because of TensorMol's flexibility, these extensions fit easily within our decomposition scheme. A general model chemistry also requires more diverse sampling of chemical space and a broader swath of the periodic table, and this work is underway in our laboratory. Accurate decomposition of bonds which always occur in pairs for example the two bonds in a terminal alkyne will benefit significantly from additional non-equilibrium geometrical data. Interested parties may download our source and train their own generalized and improved DIM-NN models.\\

\begin{acknowledgement}
The authors would like to thank Prof. Xavier Creary for valuable discussions.
\end{acknowledgement}

\begin{suppinfo}
\indent TThe Supporting Information is available free of charge via the Internet at http://pubs.acs.org/.
\end{suppinfo}

\providecommand{\latin}[1]{#1}
\makeatletter
\providecommand{\doi}
  {\begingroup\let\do\@makeother\dospecials
  \catcode`\{=1 \catcode`\}=2\doi@aux}
\providecommand{\doi@aux}[1]{\endgroup\texttt{#1}}
\makeatother
\providecommand*\mcitethebibliography{\thebibliography}
\csname @ifundefined\endcsname{endmcitethebibliography}
  {\let\endmcitethebibliography\endthebibliography}{}


\clearpage

\end{document}